\documentclass[a4paper, 11pt]{article}\usepackage[]{graphicx}\usepackage[]{color}

\usepackage{alltt}
\usepackage[left=2.5cm,right=2.5cm, top = 2.8cm, bottom = 2.8cm]{geometry}
\usepackage{todonotes}
\usepackage{amssymb}
\usepackage{amsmath}
\usepackage{url}
\usepackage{color}
\usepackage{hyperref}
\hypersetup{
colorlinks=true,
linktoc=all,
urlcolor=blue,
}

\begin{document}

\title{Evaluating epidemic forecasts in an interval format}
\author{Johannes Bracher$^{1,2}$, Evan L. Ray$^3$, Tilmann Gneiting$^{2,4}$ and Nicholas G. Reich$^3$}

\date{ \small
$^1$Karlsruhe Institute of Technology (KIT), Chair of Econometrics and Statistics \\
$^2$Heidelberg Institute for Theoretical Studies \\
$^3$University of Massachusetts, School of Public Health and Health Sciences, Department of Biostatistics and Epidemiology \\
$^4$Karlsruhe Institute of Technology (KIT), Institute for Stochastics \\ \normalsize \rule{0mm}{2.5mm} \\
    \today
}

\maketitle

\section*{Abstract}
For practical reasons, many forecasts of case, hospitalization and death counts in the context of the current COVID-19 pandemic are issued in the form of central predictive intervals at various levels. This is also the case for the forecasts collected in the \textit{COVID-19 Forecast Hub} (\url{https://covid19forecasthub.org/}). Forecast evaluation metrics like the logarithmic score, which has been applied in several infectious disease forecasting challenges, are then not available as they require full predictive distributions. This article provides an overview of how established methods for the evaluation of quantile and interval forecasts can be applied to epidemic forecasts in this format. Specifically, we discuss the computation and interpretation of the weighted interval score, which is a proper score that approximates the continuous ranked probability score. It can be interpreted as a generalization of the absolute error to probabilistic forecasts and allows for a decomposition into a measure of sharpness and penalties for over- and underprediction.

\section*{Author summary}
During the COVID-19 pandemic, model-based probabilistic forecasts of case, hospitalization and death numbers can help to improve situational awareness and guide public health interventions. The \textit{COVID-19 Forecast Hub} (\url{https://covid19forecasthub.org/}) collects such forecasts from numerous national and international groups. Systematic and statistically sound evaluation of forecasts is an important prerequisite to revise and improve models and to combine different forecasts into ensemble predictions. We provide an intuitive introduction to scoring methods, which are suitable for the interval/quantile-based format used in the Forecast Hub, and compare them to other commonly used performance measures.



\section{Introduction}

There is a growing consensus in infectious disease epidemiology that epidemic forecasts should be probabilistic in nature, i.e.\ should not only state one predicted outcome, but also quantify their own uncertainty. This is reflected in recent forecasting challenges like the US CDC \textit{FluSight} Challenge \cite{McGowan2019} and the \textit{Dengue Forecasting Project} \cite{Johansson2019}, which required participants to submit forecast distributions for binned disease incidence measures. Storing forecasts in this way enables the evaluation of standard scoring rules like the logarithmic score \cite{Gneiting2007}, which has been used in both of the aforementioned challenges. 
This approach, however, requires that a simple yet meaningful binning system can be defined and is followed by all forecasters. In acute outbreak situations like the current COVID-19 outbreak, where the range of observed outcomes varies considerably across space and time and forecasts are generated under time pressure, it may not be practically feasible to define a reasonable binning scheme. 

An alternative is to store forecasts in the form of predictive quantiles or intervals. This is the approach used in the \textit{COVID-19 Forecast Hub} \cite{UMASS2020, Ray2020}. The Forecast Hub serves to aggregate COVID-19 death and hospitalization forecasts in the United States (both national and state levels) and is the data source for the CDC COVID-19 forecasting web page \cite{CDCP2020}. Contributing teams are asked to report the predictive median and central prediction intervals with nominal levels $10\%, 20\%, \dots, 90\%, 95\%, 98\%$, meaning that the (0.01, 0.025, 0.05, 0.10, \dots, 0.95, 0.975, 0.99) quantiles of predictive distributions have to be made available. Using such a format, predictive distributions can be stored in  reasonable detail independently of the expected range of outcomes. However, suitably adapted scoring methods are required, as e.g.\ the logarithmic score cannot be evaluated based on quantiles alone. This note provides an introduction to established quantile and interval-based scoring methods \cite{Gneiting2007} with a focus on their application to epidemiological forecasts.

\section{Forecast evaluation using proper scoring rules}

\subsection{Common scores to evaluate full predictive distributions}
\label{sec:definitions}

Proper scoring rules \cite{Gneiting2007} are today the standard tools to evaluate probabilistic forecasts. Propriety is a desirable property of a score as it encourages \textit{honest} forecasting, meaning that forecasters have no incentive to report forecasts differing from their true belief about the future. We start by providing a brief overview of scores which can be applied when the full predictive distribution is available.

A widely used proper score is the \textit{logarithmic score}. In the case of a discrete set of possible outcomes $\{1, \dots, M\}$ (as is the case for counts or binned measures of disease activity), it is defined as \cite{Gneiting2007}
$$
\text{logS}(F, y) = \text{log}(p_y).
$$
Here $p_y$ is the probability assigned to the observed outcome $y$ by the forecast $F$. The log score is \textit{positively oriented}, meaning that larger values are better. A potential disadvantage of this score is that it degenerates to $-\infty$ if $p_y = 0$. In the \textit{FluSight} Challenge the score is therefore truncated at a value of $-10$ \cite{CDCP2019}; we note that when this truncation is performed, the score is no longer proper.

Until the 2018/2019 edition, a variation of the logarithmic score called the \textit{multibin} logarithmic score was used in the \textit{FluSight} Challenge. For discrete and ordered outcomes it is defined as \cite{CDCP2018}
$$
\text{MBlogS}(F, y) = \log \left( \sum_{i = -d}^d p_{y + i} \right),
$$
i.e.\ also counts probability mass within a certain tolerance range of $\pm d$ ordered categories. The goal of this score is to measure ``accuracy of practical significance'' \cite{Reich2019}. It thus offers a more accessible interpretation to practitioners, but has the disadvantage of being improper \cite{Bracher2019, Reich2019a}.

An alternative score which is considered more robust than the logarithmic score \cite{Gneiting2007a} is the continuous ranked probability score
$$
\text{CRPS}(F, y) = \int_{-\infty}^\infty \{F(x) - \mathbf{1}(x \geq y)\}^2 dx,
$$
where $F$ is interpreted as a cumulative distribution function (CDF). Note that in the case of integer-valued outcomes the CRPS simplifies to the ranked probability score, compare \cite{Czado2009} and \cite{Kolassa2016}. The CRPS represents a generalization of the absolute error to probabilistic forecasts (implying that it is negatively oriented) and has been commonly used to evaluate epidemic forecasts \cite{Held2017, Funk2019}. The CRPS does not diverge to $\infty$ even if a forecast assigns zero probability to the eventually observed outcome, making it less sensitive to occasional misguided forecasts. It depends on the application setting whether an extreme penalization of such ``missed'' forecasts is desirable or not, and in certain contexts the CRPS may seem lenient. A practical advantage, however, is that there is no need for thresholding it at an arbitrary value.

\begin{figure}[h]
\includegraphics[width=\textwidth]{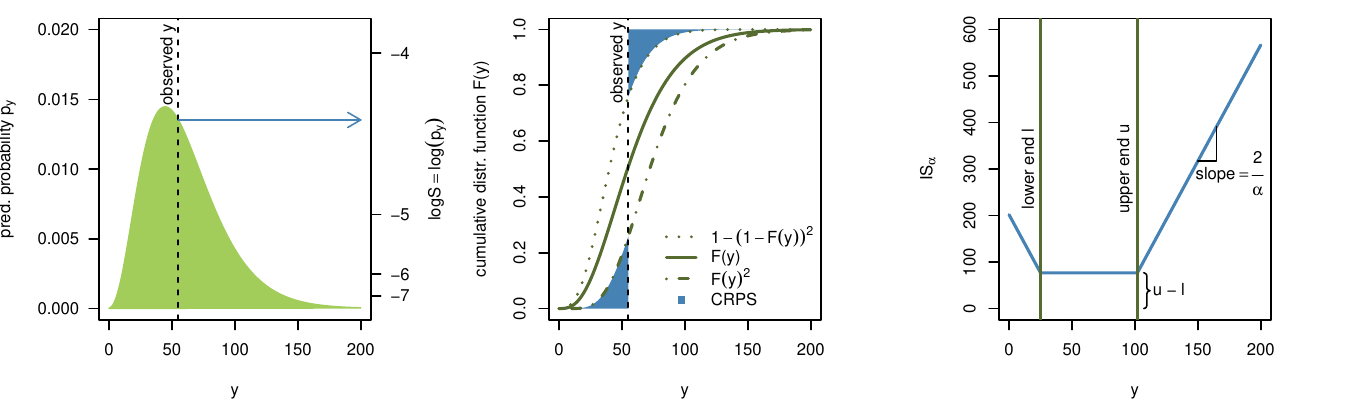} 
\caption{\textbf{Visualization of the logS, CRPS and IS.} Left: The logarithmic score only depends on the predictive probability assigned to the observed event $y$ (of which one takes the logarithm). Middle: The CRPS can be interpreted as a measure of the distance between the predictive cumulative distribution function and a vertical line at the observed value. Right: The interval score $\text{IS}_\alpha$ is a piecewise linear function which is constant inside the respective prediction interval and has slope $\pm 2/\alpha$ outside of it.}
\label{fig:definition}
\end{figure}

To facilitate an intuitive understanding of the different scores, Fig \ref{fig:definition} graphically illustrates the definitions of the logarithmic score (left) and CRPS (middle). For the CRPS note that if an observation falls far into the tails of the predictive distribution, one of the two blue areas representing the CRPS will essentially disappear, while the size of the other depends approximately linearly on the observed value $y$ (with a slope of 1). This illustrates the close link between the CRPS and the absolute error.

\subsection{Scores for forecasts provided in an interval format}
\label{subsec:interval_scores}

Both the logS and the CRPS cannot be evaluated directly if forecasts are provided in an interval format. If many intervals are provided, approximations may be feasible to some degree, but problems arise if observations fall in the tails of predictive distributions (see Discussion section). It is therefore advisable to apply scoring rules designed specifically for forecasts in a quantile/interval format. A simple proper score which requires only a central $(1 - \alpha)\times 100\%$ prediction interval (in the following: PI) is the \textit{interval score} \cite{Gneiting2007}
$$
\text{IS}_{\alpha}(F, y) = (u - l) \ + \ \frac{2}{\alpha} \times (l - y) \times \mathbf{1}(y < l) \ + \  \frac{2}{\alpha} \times (y - u) \times \mathbf{1}(y > u).
$$
Here, $\mathbf{1}$ is the indicator function, meaning that $\mathbf{1}(y < l) = 1$ if $y < l$ and 0 otherwise. The terms $l$ and $u$ denote the $\alpha/2$ and $1 - \alpha/2$ quantiles of $F$. The interval score consists of three intuitively meaningful quantities:
\begin{itemize}
\item The width $u - l$ of the central $(1 -\alpha)$ PI, which describes the sharpness of $F$.
\item A penalty term $\frac{2}{\alpha} \times (l - y) \times \mathbf{1}(y < l)$ for observations falling below the lower endpoint $l$ of the $(1 - \alpha) \times 100\%$ PI. The penalty is proportional to the distance between $y$ and the lower end $l$ of the interval, with the strength of the penalty depending on the level $\alpha$ (the higher the nominal level $(1 - \alpha) \times 100\%$ of the PI the more severe the penalty).
\item An analoguous penalty term $\frac{2}{\alpha} \times (y - u) \times \mathbf{1}(y > u)$ for observations falling above the upper end $u$ of the PI.
\end{itemize}
A graphical illustration of this definition can be found in the right panel of Fig \ref{fig:definition}. Note that the interval score has recently been used to evaluate forecasts of SARS-CoV-1 and Ebola \cite{Chowell2019} as well as SARS-CoV-2 \cite{Chowell2020}.

To provide more detailed information on the predictive distribution it is common to report not just one, but several central PIs at different levels $(1 - \alpha_1) < (1 - \alpha_2) < \dots < (1 - \alpha_K)$, along with the predictive median $m$. The latter can informally be seen as a central prediction interval at level $(1 - \alpha_0) \rightarrow 0$. To take all of these into account, a \textit{weighted interval score} can be evaluated:\footnote{Note that this definition has been modified slightly relative to previous versions of this preprint. For the previous definition
$$
\text{WIS}_{\alpha_{0:K}}(F, y) = \frac{1}{K + 1} \times \left( w_0 \times 2 \times |y  - m| \ + \ \sum_{k = 1}^K \left\{ w_k \times \text{IS}_{\alpha_k} (F, y) \right\} \right)
$$
exact equivalence to the expression in equation \eqref{eq:alternative_def} did not hold. It was therefore decided to revise the definition.}
\begin{equation} \label{eq:WIS}
\text{WIS}_{\alpha_{0:K}}(F, y) = \frac{1}{K + 1/2} \times \left( w_0 \times |y  - m| \ + \ \sum_{k = 1}^K \left\{ w_k \times \text{IS}_{\alpha_k} (F, y) \right\} \right).    
\end{equation}
This score is a special case of the more general \textit{quantile score} \cite{Gneiting2007} and it is proper for any set of non-negative (un-normalized) weights $w_0, w_1, \dots, w_K$. A natural choice is to set
\begin{equation}
w_k = \frac{\alpha_k}{2} \label{eq:w_k}
\end{equation}
with $w_0 = 1/2$, as for large $K$ and equally spaced values of $\alpha_1, \dots, \alpha_K$ (stretching over the unit interval) it can be shown that under this choice of weights
\begin{equation} \label{eq:approx_crps}
\text{WIS}_{\alpha_{0:K}}(F, y) \approx \text{CRPS}(F, y). 
\end{equation}
This follows directly from known properties of the quantile score and CRPS \cite{Laio2007, Gneiting2011}, see Appendix \ref{app:derivation}. Consequently the score can be interpreted heuristically as a measure of distance between the predictive distribution and the true observation, where the units are those of the absolute error, on the natural scale of the data. Indeed, in the case $K = 0$ where only the predictive median is used, $\text{WIS}_{\alpha_{0}}(F, y)$ is equal to the absolute error. Furthermore, the WIS and CRPS reduce to the absolute error when $F$ is a point forecast \cite{Gneiting2007}. We will use the specification \eqref{eq:w_k} of the weights in the remainder of the article, but remark that different weighting schemes may be reasonable depending on the application context.

In practice, evaluation of forecasts submitted to the \textit{COVID-19 Forecast Hub} will be done based on the predictive median and $K = 11$ prediction intervals with $\alpha_1 = 0.02, \alpha_2 = 0.05, \alpha_3 = 0.1, \dots, \alpha_{11} = 0.9$ (implying nominal coverages of $98\%, 95\%, 90\%, \dots, 10\%$). This corresponds to the quantiles teams are required to report in their submissions and implies that relative to the CRPS, slightly more emphasis is given to intervals with high nominal coverage. 

Similarly to the interval score, the weighted interval score can be decomposed into weighted sums of the widths of PIs and penalty terms, including the absolute error. These two components represent the sharpness and calibration of the forecasts, respectively, and can be used in graphical representations of obtained scores (see Section \ref{sec:graphic_display}).

Note that a score corresponding to one half of what we refer to as the WIS was used in the 2014 Global Energy Forecasting Competition \cite{Hong2016}. The score was framed as an average of pinball losses for the predictive 1st through 99th percentiles. We note in this context that with the weights $w_k$ from equation \eqref{eq:w_k} the WIS can also be expressed as
\begin{equation} \label{eq:alternative_def}
\text{WIS}_{\alpha_{0:K}}(F, y) = \frac{1}{2K + 1} \times \sum_{k = 1}^{2K + 1} 2 \times \{ \mathbf{1}(y \leq q_{\tau_k}) - \tau_k \} \times (q_{\tau_k} - y), 
\end{equation}
see Appendix \ref{app:derivation}. Here, the levels $0 < \tau_1 < \cdots < \tau_{K + 1} = 1/2 < \cdots < \tau_{2K + 1} < 1$ and the associated quantiles $q_{\tau_1}, \ldots, q_{\tau_{2K+1}}$ correspond to the median and the $2K$ quantiles defining the central PIs at levels $1 - \alpha_1, \dots, 1 - \alpha_K$.  In this paper, we preferred to motivate the score through central predictive intervals at different levels, which are a commonly used concept in epidemiology. However, when applying e.g.\ quantile regression methods for ensemble building, formulation \eqref{eq:alternative_def} may seem more natural.

\subsection{Aggregation of scores}

To compare different prediction methods systematically it is necessary to aggregate the scores they achieved over time and for various forecast targets. The natural way of aggregating proper scores is via their sum or average \cite{Gneiting2007}, as this ensures that propriety is maintained. If forecasts are made at several different horizons it is often helpful to also inspect average scores separately by horizon and assess how forecast quality deteriorates over time. Scores for longer time horizons often tend to show larger variability and can dominate average scores. This is especially true when forecasting cumulative case or death numbers, where forecast errors build up over time, and a stratified analysis can be more informative in this case.

It may be of interest to formally assess the strength of evidence that there is a difference in mean forecast skill between methods. Various tests exist to this end, the most commonly used being the Diebold-Mariano test \cite{Diebold1995}. However, this test does not have a widely accepted extension to account for dependencies between multiple forecasts made for different time-series or locations. Similar challenges arise when predictions at multiple horizons are issued at the same time. An interesting strategy in this context is to treat these predictions as a path forecast and assess them jointly \cite{Pinson2012, Golestaneh2016}.  Generally, theoretically principled methods for multivariate forecast evaluation exist \cite{Gneiting2008} and have found applications in disease forecasting \cite{Held2017}.

A topic closely related to path forecasting is forecasting of more qualitative or longer-term characteristics of an epidemic curve. For instance, in the FluSight challenges \cite{McGowan2019} forecasts for the timing and strength of seasonal peaks have been assessed. While such targets can be of great interest from a public health perspective, it is not always obvious how to define them for an emerging rather than seasonal disease. A possibility would be to consider maximum weekly incidences over a gliding time window. This could provide additional information on the peak healthcare demand expected over a given time period, an aspect which is often not reflected well in independent week-wise forecasts \cite{Juul2020}.

\section{Qualitative comparison for different scores}
\label{sec:qualitative_comparison}

We now compare various scores using simple examples, covering scores for point predictions, prediction intervals and full predictive distributions.

\subsection{Illustration for an integer-valued outcome}

\begin{figure}[h]
\includegraphics[width=\textwidth]{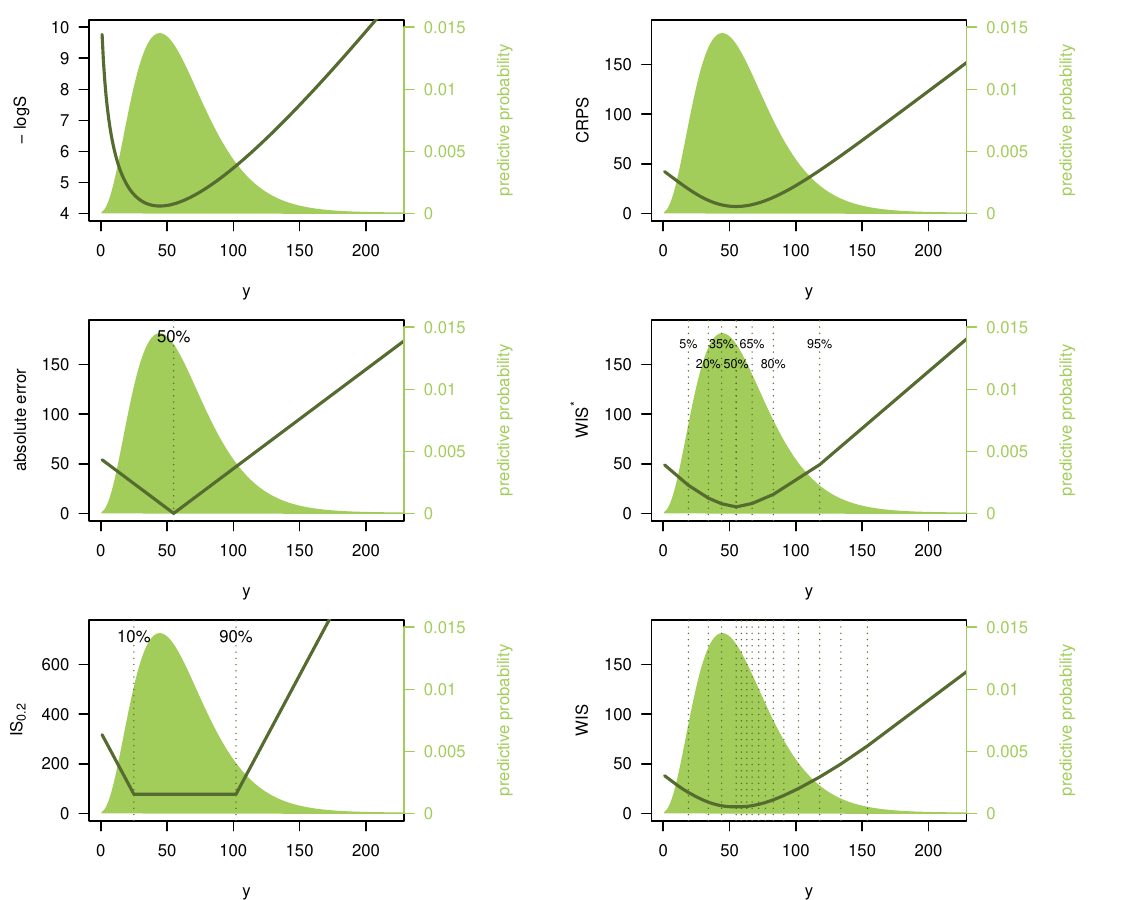} 
\caption{\textbf{Illustration of different scoring rules.} Logarithmic score, absolute error, interval score (with $\alpha = 0.2$), CRPS and two versions of the weighted interval score. These are denoted by $\text{WIS}^*$ (with $K = 3$, $\alpha_1 = 0.1, \alpha_2 = 0.4, \alpha_3 = 0.7$) and WIS ($K = 11$, $\alpha_1 = 0.02, \alpha_2 = 0.05, \alpha_3 = 0.1, \dots, \alpha_{11} = 0.9$). Scores are shown as a function of the observed value $y$. The predictive distribution $F$ is negative binomial with expectation 60 and size 4. Note that the top left panel shows the negative logS, i.e.~$ -\text{logS}$, which like the other scores is negatively oriented (smaller values are better).}
\label{fig:ls_vs_WIS}
\end{figure}

Fig \ref{fig:ls_vs_WIS} illustrates the behaviour of five different scores for a negative binomial predictive distribution $F$ with expectation $\mu_F = 60$ and size parameter $\psi_F = 4$ (standard deviation $\approx 31.0$).  We consider the logarithmic score, absolute error, interval score with $\alpha = 0.2$ ($\text{IS}_{0.2}$), CRPS, and two versions of the weighted interval score. Firstly, we consider a score with $K = 3$ and $\alpha_1 = 0.1, \alpha_2 = 0.4, \alpha_3 = 0.7$, which we denote by $\text{WIS}^*$. Secondly, we consider a more detailed score with $K = 11$ and $\alpha_1 = 0.02, \alpha_2 = 0.05, \alpha_3 = 0.1, \dots, \alpha_{11} = 0.9$, denoted by WIS (as this is the version used in the \textit{COVID-19 Forecast Hub} we will focus on it in the remainder of the article). The resulting scores are shown as a function of the observed value $y$. Qualitatively all curves look similar. However, some differences can be observed. The best (lowest) negative logS is achieved if the observation $y$ coincides with the predictive mode. For the interval-based scores, AE and CRPS, the best value results if $y$ equals the median (for the $\text{IS}_{0.2}$ in the middle right panel there is a plateau as it does not distinguish between values falling into the 80\% PI). The negative logS curve is more smooth and increases the more steeply the further away the observed $y$ is from the predictive mode. The curve shows some asymmetry, which is absent or less pronounced in the other plots. The IS and WIS curves are piecewise linear. The WIS has a more modest slope closer to the median and a more pronounced one towards the tails (approaching $-1$ and 1 in the left and right tail, respectively). Both versions of the WIS represent a good approximation to the CRPS. For the more detailed version with 11 intervals plus the absolute error, slight differences to the CRPS can only be seen in the extreme upper tail. When comparing the  CRPS and $\text{WIS}^*$/WIS scores to the absolute error, it can be seen that the latter are larger in the immediate surroundings of the median (and always greater than zero), but lower towards the tails. This is because they also take into account the uncertainty in the forecast distribution. 

\subsection{Differing behaviour if agreement between predictions and observations is poor}
\label{subsec:differing_behaviour}

\begin{figure}[t]
\includegraphics[width=1\textwidth]{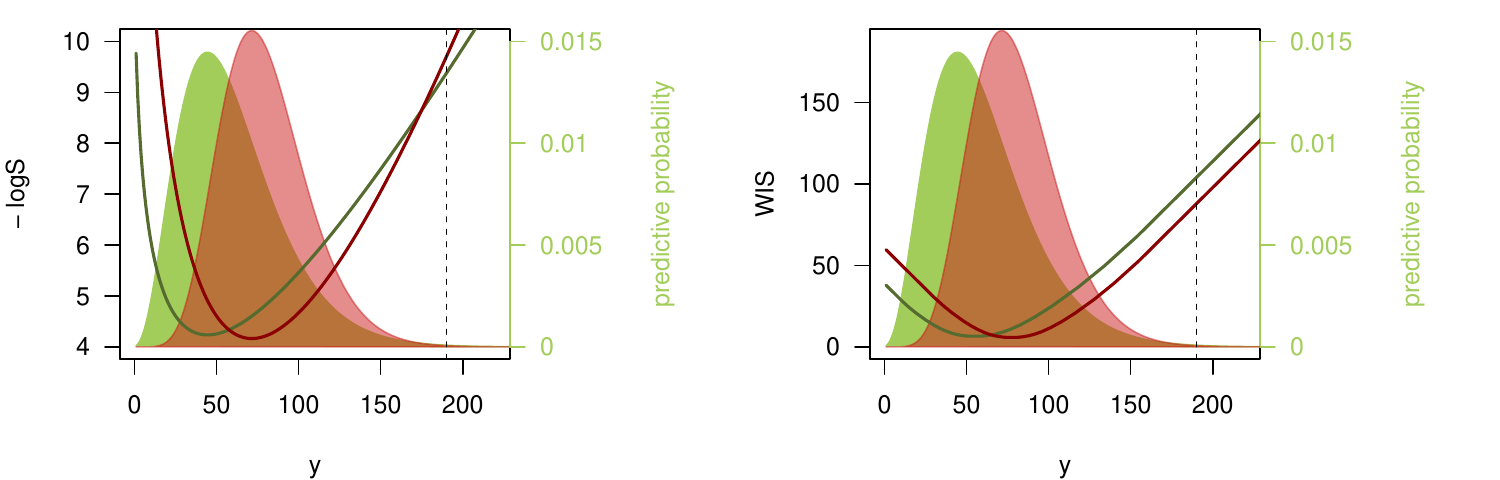} 
\caption{\textbf{Disagreement between logarithmic score and WIS.} Negative logarithmic score and weighted interval score (with $\alpha_1 = 0.02, \alpha_2 = 0.05, \alpha_3 = 0.1, \dots, \alpha_{11} = 0.9$) as a function of the observed value $y$. The predictive distributions $F$ (green) and $G$ (red) are negative binomials with expectations $\mu_F = 60, \mu_G = 80$ and sizes $\psi_F = 4, \psi_G = 10$. The black dashed line shows $y = 190$ as discussed in the text.}
\label{fig:tails}
\end{figure}

Qualitative differences between the logarithmic and interval-based scores occur predominantly if observations fall into the tails of predictive distributions. 
We illustrate this with a second example. Consider two negative binomial forecasts: $F$ with expectation 60 and size 4 (standard deviation $\approx 31$) as before, and $G$ with expectation 80, and size 10 (standard deviation $\approx 26.8)$. $G$ thus has higher expectation than $F$ and is sharper. If we now observe $y = 190$, i.e.\ a count considerably higher than suggested by either $F$ or $G$, the two scores yield different results, as illustrated in Fig \ref{fig:tails}.
\begin{itemize}
\item The logS favours $F$ over $G$, as the former is more dispersed and has slightly heavier tails. Therefore $y = 190$ is considered somewhat more ``plausible'' under $F$ than under $G$ ($\text{logS}(F, 190) = -9.37$, $\text{logS}(G, 190) = -9.69$).
\item The WIS (with $K = 11$ as in the previous section), on the other hand, favours $G$ as its quantiles are generally closer to the observed value $y$ ($\text{WIS}(F, 190) = 103.9$, $\text{WIS}(G, 190) = 87.8$).
\end{itemize}

This behaviour of the WIS is referred to as \textit{sensitivity to distance} \cite{Gneiting2007}. In contrast, the logS is a \textit{local} score which ignores distance. Winkler \cite{Winkler1996} argues that local scoring rules can be more suitable for inferential problems, while sensitivity to distance is reasonable in many decision making settings. In the public health context, say a prediction of hospital bed need on a certain day in the future, it could be argued that for $y = 190$ the forecast $G$ was indeed more useful than $F$. While a pessimistic scenario under $G$ (defined as the 95\% quantile of the predictive distribution) implies 128 beds needed and thus fell considerably short of $y = 190$, it still suggested more adequate need for preparation than $F$, which has a 95\% quantile of 118.

We argue that poor agreement between forecasts and observations is more likely to occur for COVID-19 deaths than e.g.\ for seasonal ILI (influenza-like illness) intensity, which due to larger amounts of historical data is more predictable. Sensitivity to distance then leads to more robust scoring with respect to decision making, without the need to truncate at an arbitrary value (as required for the log score). While these are pragmatic statistical considerations, it could be argued that the choice of scoring rule should depend on the cost of different types of errors. If the cost of single misguided forecasts is very high, a conservative evaluation approach using the logarithmic score may be more appropriate. This question is obviously linked to the preferences and priorities of forecasts recipients, in our case public health officials and the general public. Studying these preferences in more detail is an interesting avenue for further research.

\section{Application to FluSight forecasts}

In this section some additional practical aspects are discussed using historical forecasts from the 2016/2017 edition of the \textit{FluSight} Challenge. Note that these were originally reported in a binned format, but for illustration we translated them to a quantile format for some of the below examples.

\subsection{An easily interpretable graphic display of the WIS}
\label{sec:graphic_display}

The decomposition of the WIS into the average width of PIs and average penalty for observations outside the various PIs (see Section \ref{sec:definitions}) enables an intuitive graphical display to compare different forecasts and understand why one forecast outperforms another. Distinguishing also between penalties for over- and underprediction can be informative about systematic biases or asymmetries. Note that decompositions of quantile or interval scores for visualization purposes have been suggested before, see e.g.\ \cite{Bentzien2014}.

\begin{figure}[h!]
\center
\includegraphics[width = 0.9\textwidth]{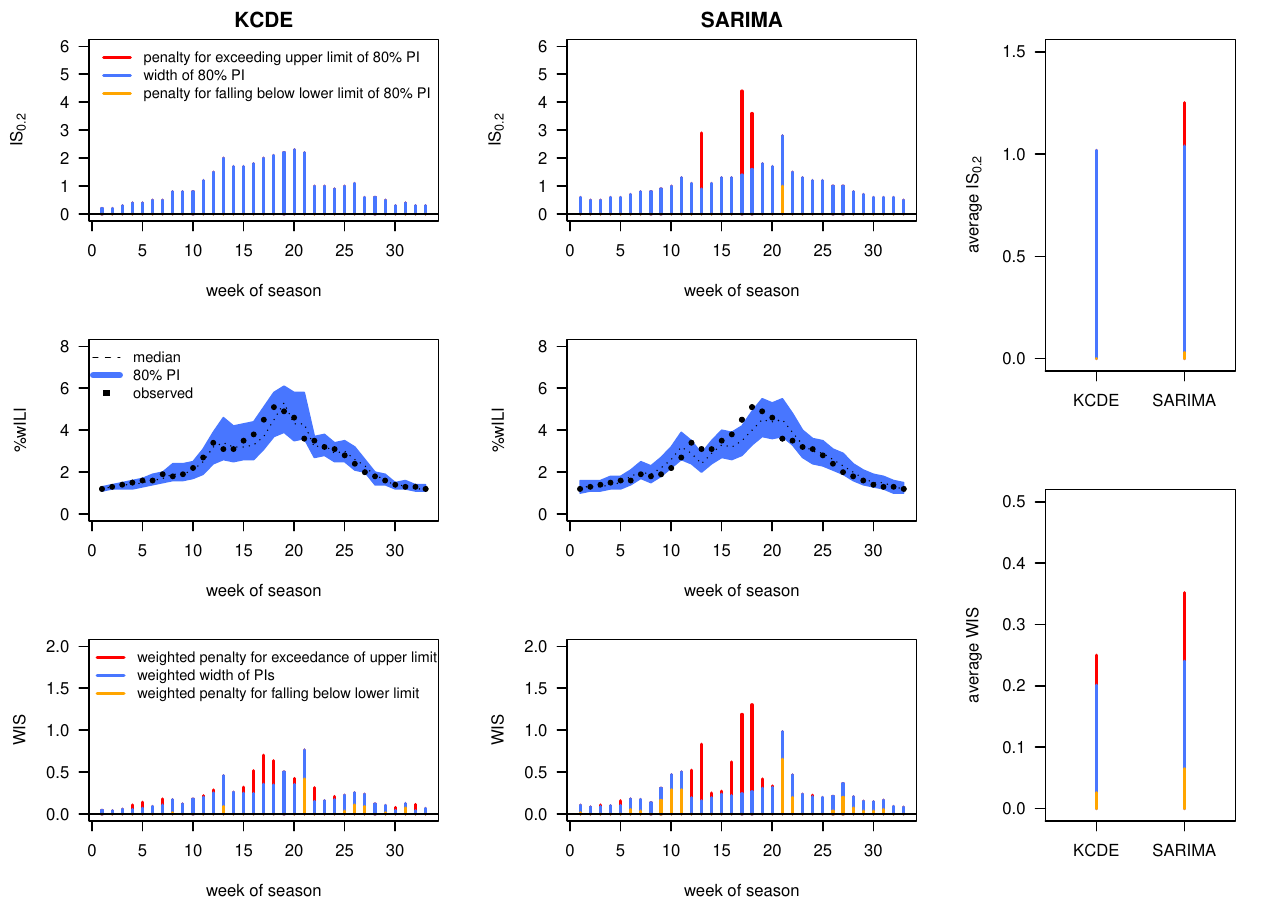}
\caption{\textbf{Interval and weighted interval score applied to FluSight forecasts.} Comparison of one-week-ahead forecasts by KCDE and SARIMA over the course of the 2016/2017 \textit{FluSight} season. The top row shows the interval score with $\alpha = 0.2$, the bottom row the weighted interval score with $\alpha_1 = 0.02$, $\alpha_2 = 0.05$, $\alpha_3 = 0.1$, $\dots$, $\alpha_{11} = 0.9$. The panels at the right show mean scores over the course of the season. All bars are decomposed into the contribution of interval widths (i.e. a measure of sharpness; blue) and penalties for over- and underprediction (orange and red, respectively). Note that the absolute values of the two scores are not directly comparable as the WIS involves re-scaling of the included interval scores.}
\label{fig:kcde_sarima}
\end{figure}

Fig \ref{fig:kcde_sarima} shows a comparison of the $\text{IS}_{0.2}$ and $\text{WIS}$ (with $K = 11$ as before) 
obtained for one-week-ahead forecasts by the KCDE and SARIMA models during the 2016/2017 \textit{FluSight} Challenge, \cite{Reich2019, Ray2017}, using data obtained from \cite{FluSight2020}. It can be seen that, while KCDE and SARIMA issued forecasts of similar sharpness (average widths of PIs, blue bars), SARIMA is more strongly penalized for PIs not covering the observations (orange and red bars). Broken down to a single number, the bottom right panel shows that predictions from KCDE and SARIMA were on average off by 0.25 and 0.35 percentage points, respectively (after taking into account the uncertainty implied by their predictions). Both methods are somewhat conservative, with 80\% PIs covering 88\% (SARIMA) and 100\% of the observations (KCDE). When comparing the plots for $\text{IS}_{0.2}$ and $\text{WIS}$, it can be seen  that the former strongly punishes larger discrepancies between forecasts and observations while ignoring smaller differences. The latter translates discrepancies to penalties in a smoother fashion, as could already be seen in Fig \ref{fig:ls_vs_WIS}.

\subsection{Visually assessing calibration}

\begin{figure}[h!]
\center
\includegraphics[scale=0.55]{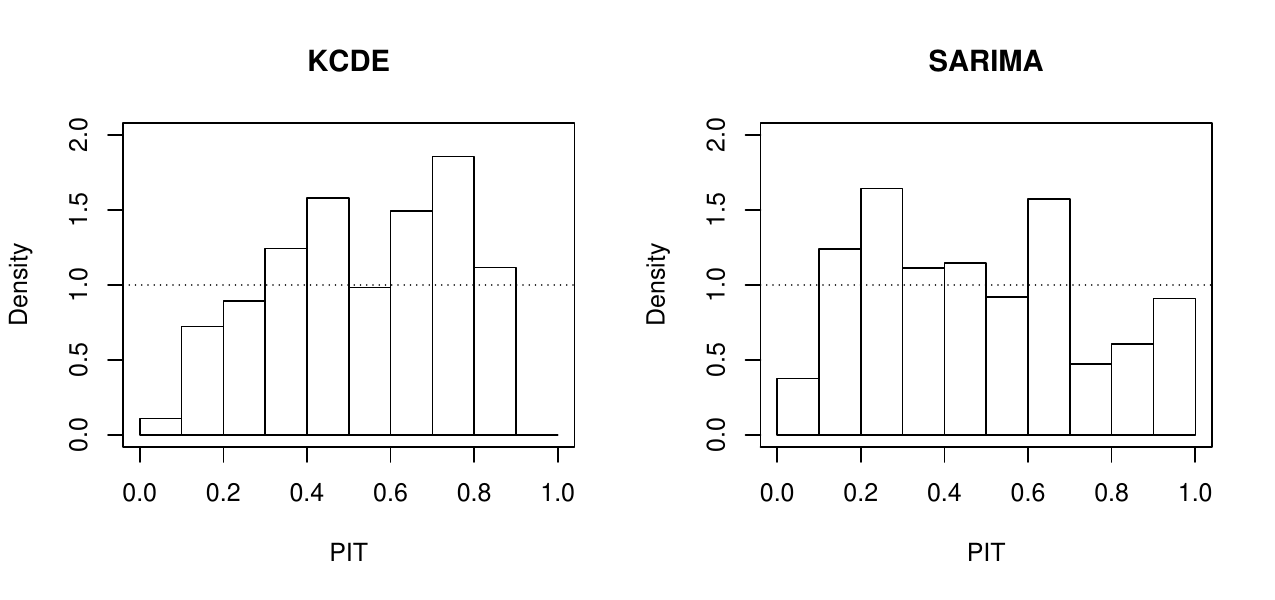}
\caption{\textbf{PIT histograms.} PIT histograms for one-week-ahead forecasts from the KCDE and SARIMA models, 2016--2017 \textit{FluSight} season. Note that to account for the discreteness of the binned distribution we employed the non-randomized correction suggested in \cite{Czado2009}.}
\label{fig:PIT}
\end{figure}

While the middle row of Fig \ref{fig:kcde_sarima} provides a good intuition of the sharpness of different forecasts, different visual tools exist to assess their calibration. A commonly used approach is via \textit{probability integral transform} (PIT) histograms \cite{Dawid1984, Czado2009}, see also \cite{Held2017} for adaptations to count data. These show the empirical distribution of
$$
\text{PIT}(F, y_{\text{obs}}) = F(y_{\text{obs}})
$$
across different forecasts from the same model. Here $F$ represents the predictive cumulative distribution function. For a calibrated forecast the PIT histogram should be approximately uniform, and deviations from uniformity indicate bias or problems with the dispersion of forecasts. E.g., L-shaped PIT histograms indicate a downward bias of forecasts, while a J shape indicates an upward bias. Under- and overdispersed forecasts lead to U and inverse-U-shaped PIT histograms, respectively. Fig \ref{fig:PIT} shows PIT histograms for one-week-ahead forecasts from KCDE and SARIMA. While no apparent biases can be seen, both models seem to produce forecasts with too high dispersion. This is especially visible for KCDE, for which hardly any realizations fell into the two extreme deciles of the respective forecast distributions.

The exact PIT values cannot be computed if forecasts are reported in a quantile format, but if sufficiently many quantiles are available one can evaluate in which decile or ventile they fall. This is sufficient to represent them graphically in a histogram with the respective number of bins. A technical problem arises if an observation is exactly equal to one or several of the reported quantiles (which can happen especially in low count settings). The corrections for discreteness suggested by \cite{Czado2009} cannot be applied in this case and there does not seem to be a standard approach for this. A practical strategy is to split up such a count between the neighbouring bins of the histogram (assigning 1/2 to each bin if the realized value coincides with one reported quantile, 1/4, 1/2, 1/4 if it coincides with two, 1/6, 1/3, 1/3, 1/6 if it coincides with three, and so on).

\subsection{Empirical agreement between different scores}

\begin{figure}[h!]
\center
\includegraphics[width = 0.82\textwidth]{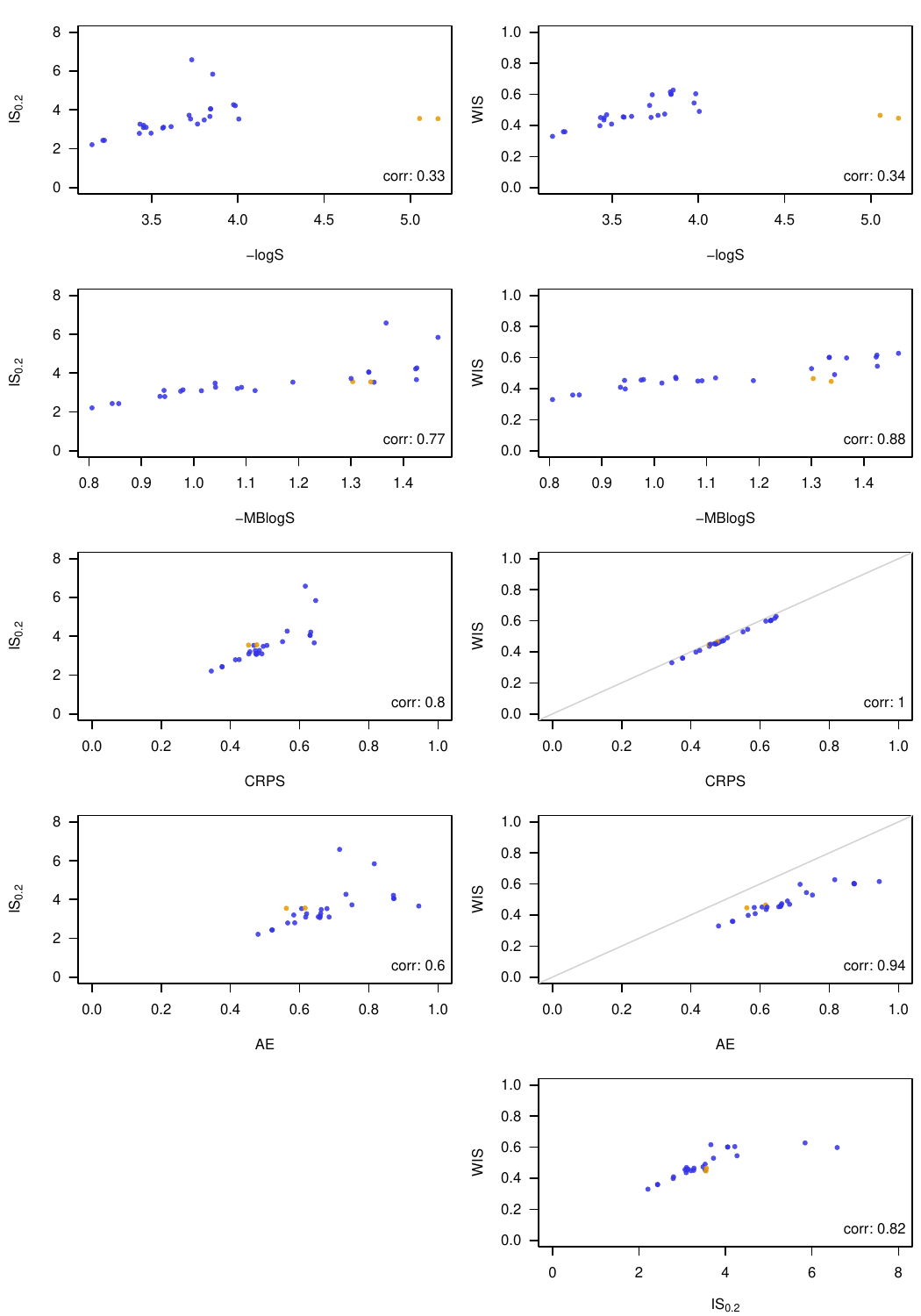}
\caption{\textbf{Comparison of 26 models participating in the 2016/2017 \textit{FluSight} Challenge under different scoring rules.} Shown are mean scores averaged over one through four week ahead forecasts, different geographical levels, weeks and forecast horizons. Compared scores: negative logarithmic score and multibin logarithmic score, continuous ranked probability score, interval score ($\alpha = 0.2$), weighted interval score with $K = 11$. Plots comparing the WIS to CRPS and AE, respectively, also show the diagonal in grey as these three scores operate on the same scale. All shown scores are negatively oriented. The models \texttt{FluOutlook\_Mech} and \texttt{FluOutlook\_MechAug} are highlighted in orange as they rank very differently under different scores.}
\label{fig:flusight}
\end{figure}

To explore the agreement between different scores, we applied several of them to one- through four-week-ahead forecasts from the 2016/2017 edition of the \textit{FluSight} Challenge. We compare the negative logarithmic score, the negative multibin logarithmic score with a tolerance of 0.5 percentage points (both with truncation at $-10$), the CRPS, the absolute error of the median, the interval score with $\alpha = 0.2$ and the weighted interval score (with $K = 11$ and $\alpha_1 = 0.02, \alpha_2 = 0.05, \alpha_3 = 0.1, \dots, \alpha_{11} = 0.9$ as in the previous sections). To evaluate the CRPS and interval scores we simply identified each bin with its central value to which a point mass was assigned. Fig \ref{fig:flusight} shows scatterplots of mean scores achieved by 26 models (averaged over weeks, forecast horizons and geographical levels; the na\"ive uniform model was removed as it performs clearly worst under almost all metrics).

As expected, the three interval-based scores correlate more strongly with the CRPS and the absolute error than with the logarithmic score. Agreement between the $\text{WIS}$ and CRPS is almost perfect, meaning that in this example the approximation \eqref{eq:approx_crps} works quite well based on the 23 available quantiles. Agreement between the interval-based score and the logS is mediocre, in part because the models  \texttt{FluOutlook\_Mech} and \texttt{FluOutlook\_MechAug} receive comparatively good interval-based scores (as well as CRPS, absolute errors and even MBlogS), but exceptionally poor logS. The reason is that while having a rather accurate central tendency, they are too sharp with tails that are too light. This is sanctioned severely by the logarithmic score, but much less so by the other scores (this is related to the discussion in Section \ref{subsec:differing_behaviour}). The $\text{WIS}$ score (and thus also the CRPS) shows remarkably good agreement with the MBlogS, indicating that distance-sensitive scores may be able to formalize the idea of a score which is slightly more ``generous" than the logS while maintaining propriety. Interestingly, all scores agree that the three best models are \texttt{LANL\_DBMplus}, \texttt{Protea\_Cheetah} and \texttt{Protea\_Springbok}.

\section{A brief remark on evaluating point forecasts}

While the main focus of this note is on the evaluation of forecast intervals, we also briefly address how point forecasts submitted to the COVID-19 Forecast Hub will be evaluated. As in the \textit{FluSight} Challenge \cite{CDCP2019}, the absolute error (AE) will be applied. This implies that teams should report the predictive median as a point forecast \cite{Gneiting2011a}. Using the absolute error in combination with WIS is appealing as both can be reported on the same scale (that of the observations). Indeed, as mentioned before, the absolute error is the same as the WIS (and CPRS) of a distribution putting all probability mass on the point forecast.

The absolute error, when averaged across time and space, is dominated by forecasts from larger states and weeks with high activity (this also holds true for the CRPS and WIS). One may thus be tempted to use a relative measure of error instead, such as the mean absolute percentage error (MAPE). We argue, however, that emphasizing forecasts of targets with higher expected values is meaningful. For instance, there should be a larger penalty for forecasting 200 deaths if 400 are eventually observed than for forecasting 2 deaths if 4 are observed. Relative measures like the MAPE would treat both the same. Moreover, the MAPE does not encourage reporting predictive medians nor means, but rather obscure and difficult to interpret types of point forecasts \cite{Gneiting2011a, Kolassa2016}. It should therefore be used with caution.

\section{Discussion}

In this paper we have provided a practical and hopefully intuitive introduction on the evaluation of epidemic forecasts provided in an interval or quantile format. It is worth emphasizing that the concepts underlying the suggested procedure are by no means new or experimental. Indeed, they can be traced back to \cite{Dunsmore1968} and \cite{Winkler1972}. As mentioned before, a special case of the WIS was used in the 2014 \textit{Global Energy Forecasting Competition} \cite{Hong2016}. A scaled version of the interval score was used in the 2018 \textit{M4} forecasting competition \cite{Makridakis2020}. The ongoing \textit{M5} competition uses the so-called \textit{weighted scaled pinball loss} (WSPL), which can be seen as a scaled version of the WIS based on the predictive median and 50\%, 67\%, 95\% and 99\% PIs \cite{MOFC2020}.

Note that we restrict attention to the case of \textit{central} prediction intervals, so that each prediction interval is clearly associated with two quantiles. The evaluation of prediction intervals which are not restricted to be central is conceptually challenging \cite{Askanazi2018, Brehmer2020}, and we refrain from adding this complexity.

The method advocated in this note corresponds to an approximate CRPS computed from prediction intervals at various levels. A natural question is whether such an approximation would also be feasible for the logarithmic score, leading to an evaluation metric closer to that from the \textit{FluSight} Challenge. We see two principal difficulties with such an approach. Firstly, some sort of interpolation method would be needed to obtain an approximate density or probability mass function within the provided intervals. While the best way to do this is not obvious, a pragmatic solution could likely be found. A second problem, however, would remain: For observations outside of the prediction interval with the highest nominal coverage (98\% for the \textit{COVID-19 Forecast Hub}) there is no easily justifiable way of approximating the logarithmic score, as the analyst necessarily has to make strong assumptions on the tail behaviour of the forecast. As such poor forecasts typically have a strong impact on the average log score, they cannot be neglected. And given that forecasts are often evaluated for many locations (e.g., over 50 US states and territories), even for a perfectly calibrated model there will on average be one such observation falling in the far tail of a predictive distribution every week. One could think about including even more extreme quantiles to remedy this, but forecasters may not be comfortable issuing these and the conceptual problem would remain. This is linked to the general problem of low robustness of the logarithmic score. We therefore argue that especially in contexts with low predictability such as the current COVID-19 pandemic, distance-sensitive scores like the CRPS or WIS are an attractive option.

\appendix

\section{Relationship between quantile score, interval score and CRPS}
 \label{app:derivation}

The standard piecewise linear quantile score \cite{Gneiting2007, Gneiting2011a} for the level $\tau$ is defined as
$$
\text{QS}_\tau(F, y) = 2 \times \{\mathbf{1}(y \leq q_\tau) - \tau\} \times (q_\tau - y),
$$
where $q_\tau$ is the $\tau$ quantile of the forecast $F$ and $y$ is the observed outcome. It can be shown by some re-ordering of terms that the interval score of a central ($1 - \alpha$) PI can be computed from the quantile scores at levels $\alpha/2$ and $1 - \alpha/2$ as

\begin{equation} \label{eq:is_qs}
\text{IS}_\alpha(F, Y) = \frac{\text{QS}_{\alpha/2}(F, y) + \text{QS}_{1 - \alpha/2}(F, y)}{\alpha}.
\end{equation}

Interestingly, this is the only available proper interval score that is invariant under translation \cite[Theorem 4]{Brehmer2020}, so that for a prediction horizon of one time unit, evaluations in terms of incident counts yield the same results as evaluations in terms of cumulative counts.

Moreover it is known \cite{Laio2007, Gneiting2011} that

\begin{align}
\text{CRPS}(F, y) & = \int_0^1 \text{QS}_\tau(F, y) \, \text{d} \tau, \nonumber \\
 & \approx \frac{1}{2K + 1} \times \sum_{k = 1}^{2K + 1} \text{QS}_{\tau_k}(F, y), \nonumber \\
 & = \frac{1}{2K + 1} \times \sum_{k = 1}^{2K + 1} 2 \times \{\mathbf{1}(y \leq q_{\tau_k}) - \tau_k\} \times (q_{\tau_k} - y), \label{eq:crps_qs}
 \end{align}
 
with a large number of (approximately) equally spaced levels $\tau_k$ stretching the unit interval such that $\tau_1 < \cdots < \tau_{K + 1} = 1/2 < \cdots < \tau_{2K+1}$. Note that expression \eqref{eq:crps_qs} is the same as the alternative expression \eqref{eq:alternative_def} for the WIS from the main text, where $\tau_k = \alpha_k/2$ and $\tau_{2K+2-k} = 1 - \alpha_k/2$ for $k =1, \ldots, K$. 

Indeed, starting from the original definition of the WIS in equation \eqref{eq:WIS} with weights $w_0 = 1/2$ and $w_k = \alpha_k/2$ for $k = 1, \ldots, K$ as in equation \eqref{eq:w_k} from the main text, using equation \eqref{eq:is_qs}, and noting that $\tau_{K + 1} = 1/2$ and $q_{\tau_{K + 1}} = m$ is the median, we see that

\begin{align*}
\text{WIS}_{\alpha_{0:K}}(F, y) & = \frac{1}{2K + 1} \times \left( |y - m| \ + \ \sum_{k = 1}^K \alpha_k \times \text{IS}_{\alpha_k}(F, y) \right) \\
 & = \frac{1}{2K + 1} \left( \text{QS}_{\tau_{K + 1}}(F, y) + \sum_{k = 1}^K \left\{ \text{QS}_{\tau_{k}}(F, y) + \text{QS}_{\tau_{2K + 2 - k}}(F, y) \right\} \right) \\
 & = \frac{1}{2K + 1} \times \sum_{k = 1}^{2K + 1} 2 \times \{\mathbf{1}(y \leq q_{\tau_k}) - \tau_k\} \times (q_{\tau_k} - y).
\end{align*}

As noted in Section \ref{subsec:interval_scores}, the $\tau_k$ we use in practice are not equally spaced in the tails (due to the addition of the quantiles at levels 0.01, 0.025, 0.975, and 0.99 forming the 95\% and 98\% prediction intervals). Relative to the CRPS, we thus put slightly more weight on the tails.

\section*{Reproducibility}

Code to reproduce Fig 1--6 has been made available at\\ \url{https://github.com/reichlab/proper-scores-comparison}. All data used in this paper have been taken from the public \texttt{cdc-flusight-ensemble} repository \cite{FluSight2020}.

\section*{Acknowledgements}

We thank Ryan Tibshirani and Sebastian Funk for their insightful comments. The work of Johannes Bracher was supported by the Helmholtz Foundation via the SIMCARD Information \& Data Science Pilot Project. Tilmann Gneiting is grateful for support by the Klaus Tschira Foundation.  Evan L. Ray and Nicholas G. Reich were supported by the US Centers for Disease Control and Prevention (1U01IP001122). The content is solely the responsibility of the authors and does not necessarily represent the official views of the CDC.

\end{document}